\documentclass[aps,pre,twocolumn,groupedaddress,showpacs]{revtex4}
   \input epsf
\input rotate

\usepackage{graphicx}

\begin{document}

\title{Experimental evidence of flow destabilization in a 2D bidisperse foam.}

\author{I. Cantat}
\email[]{isabelle.cantat@univ-rennes1.fr}
\author{C. Poloni}
\author{R. Delannay}
\affiliation{GMCM, UMR 6626, Universit\'e de Rennes (CNRS),
                   263, av. du G\'en\'eral Leclerc
                      35042 Rennes Cedex, France }

\date{\today}

\begin{abstract}
Liquid foam flows in a Hele-Shaw cell were investigated. The plug flow obtained for a monodisperse foam  is
 strongly  perturbed in the presence of bubbles whose size is larger than the average bubble size by an order 
of magnitude at least. The large bubbles migrate faster than the mean flow
above a velocity threshold
 which depends on its size. We evidence
experimentally this new  instability  and, in case of a single large bubble, we
compare the large bubble velocity with the prediction deduced from 
scaling arguments. In case of a bidisperse 
foam, an attractive interaction between large bubbles induces segregation and the large bubbles 
organize themselves in columns oriented along the flow.
These results allow to identify the main ingredients governing 2D polydisperse foam flows.
\end{abstract}


\maketitle
 \section{Introduction}

In the context of complex and structured fluids, 2D foam rheology gives rise to a revived
interest. In these materials, the subtle interplay between the elastic, plastic and viscous behaviors 
induces complex 
rheological properties.
The similarities between fluid foam,  pastes, clays or slurries arise from 
the organisation at small scale of a disordered structure, that governs the macroscopic behavior.
In case of foam, this small scale is the millimetric bubble scale, which is much easier
 to observe than the molecular scale usually involved. The structure determination is especially convenient 
in 2D foam which is thus often used as a model system for complex fluids \cite{fortes99,debregeas01,
lauridsen02,asipauskas03}. 

  We developed an experimental set-up enabling foams to flow in a dissipative regime in a large 2D channel.
 The foam is confined between two horizontal glass plates separated by a small gap
allowing a sole bubble layer to form. The whole films network, as well as the velocity fields, can thus be determined from
images taken from above, where bubbles appear as polygons of various areas.  Each vertical film
 touches both plates along lines called {\it Plateau borders}. When sliding on the plates, they induce
a viscous dissipation that becomes important even at relatively small velocities \cite{hirasaki85, cantat04}.
When a foam is pushed in the cell, it is thus submitted to a viscous force depending on its velocity and on the density of Plateau
borders, {\it ie} on the bubble sizes. If the foam is polydisperse, the largest bubbles experience
 a smaller drag (per unit foam surface, in a 2D language) than the smallest ones and
 tend therefore to move faster than the mean flow. 
The resulting flow is difficult to quantify in the general case of a polydisperse foam.
As a first step, we investigated the case of a single large bubble embedded in a sea of much smaller
bubbles and the case of a bidisperse foam with a minority of large bubbles. These large bubbles, that play
 the role of defects (or holes) in the foam, can be produced {\it a posteriori} with a controlled
size and position by vaporisation of liquid films with a laser. 

As in a viscous instability of an
interface, the pressure gradient in the foam is higher in-front of
the hole than to the sides. A similar hole in a simple viscous liquid
would move ahead for this reason. In a foam,  the shear stress created by the
different pressure gradients must exceed a yield stress. A
threshold velocity is thus set by a competition between surface tension and dissipative
forces. 
Only above this velocity, the large bubbles begin to migrate through the smaller bubbles. This migration 
 does not imply that films break : the process is based on elementary neighbors exchanges involving
four  bubbles called
{\it T1} events \cite{weaire}(see Fig. \ref{T1}). 
The flow destabilization presented in this paper is based on wall effects and should thus be more important
at smaller scale and especially in the domain of microfluidics. 

This enhanced  role of the dissipation
 in 2D foam flows in presence of large bubbles has been predicted numerically but 
 is experimentally evidenced here for the first time. 
We present the experimental set-up in the section 2 and the qualitative flow behavior 
in section 3. Then, in section 4, we briefly recall the theoretical predictions previously obtained in \cite{cantat03a,cantat05a}. The 
section 5 is devoted to the relative velocity of a single large bubble in a monodisperse foam flow, which is the main result of this paper. Finally, in section 6, 
 we show that size segregation occurs in the case of several large bubbles and that these bubbles organize themselves in 
columns oriented in the direction of the flow.

\section{Experimental set-up}

\begin{figure}
\centering
\includegraphics[width=6.5cm]{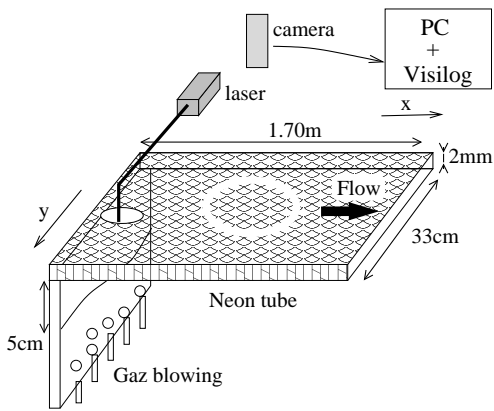}
\caption{\it Experimental set-up.}
\label{setup}
\end{figure}

Foams were prepared from a solution of SDS (3g/L), dodecanol (0.01g/L) and glycerol
(0.05 L/L) in ultra pure water. As-prepared solutions were used within a period of 24h.
The surface tension is $\gamma = 31 \,  10^{-3} N/m$ and the bulk viscosity
is $\eta = 1.1\,  10^{-3} Pa\,  s$.
The cell flow is made of two large horizontal glass plates ($l= 35cm \times L=170 cm$) separated by a 
gap $h=2mm$ (fig. 1). It is connected upstream to a vertical production cell, of the same width and thickness, containing the
 foaming solution.
The gap thickness is ensured by a steel piece, on which the glass
plates are simply clamped. 
 The foam is produced by blowing nitrogen continuously at a controlled flow rate through 4 equally spaced needles of diameter 1mm. It drains
 in the vertical cell until it reaches the main cell (fig.\ref{setup}). The present data were collected with a
 drainage height of 5 cm. The resulting liquid fraction is difficult to estimate but it corresponds to a very dry foam.

Depending on the gas flow rate we impose, the turbulence in the foaming solution is modified and
 yields various polydispersities.
We worked with almost monodisperse
foams with a typical bubble volume of $V = 5 \,  10^{-2} mL$ (which gives a typical diameter of $d=5 mm \pm 1$ for the polygons), obtained at the small flow rate of 1.5 mL/s per injector.
The large bubble is produced by a YAG laser (wave length 1064 nm, energy per pulse 20 mJ, pulse duration 5 ns). The
 beam is successively focused on each desired film with an orientable lens, until a bubble 10 to 400 times larger in volume  is 
 produced.
Once the foam and the defect are produced at low flux, the foam flow is accelerated at the
 required velocity by increasing the gas flow rate. Velocities were varied between $0.3 cm/s$ and $10cm/s$.
 The structure of the foam produced at this stage may be highly disordered but it does not influence
 the flow observed downstream.

\begin{figure}
\centering
\includegraphics[width=3.5cm]{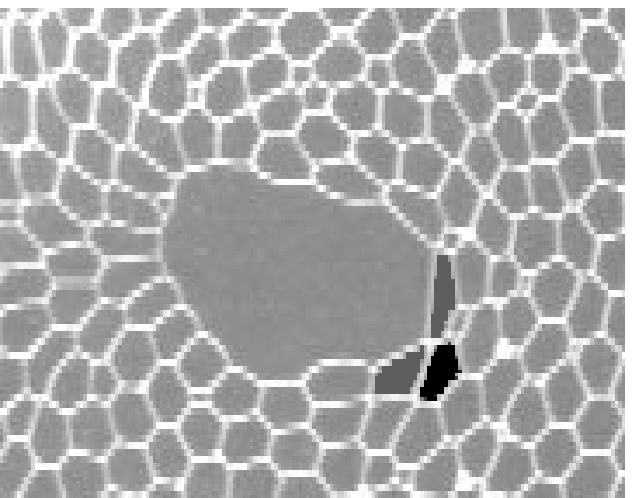}\hskip 0.2cm
\includegraphics[width=3.5cm]{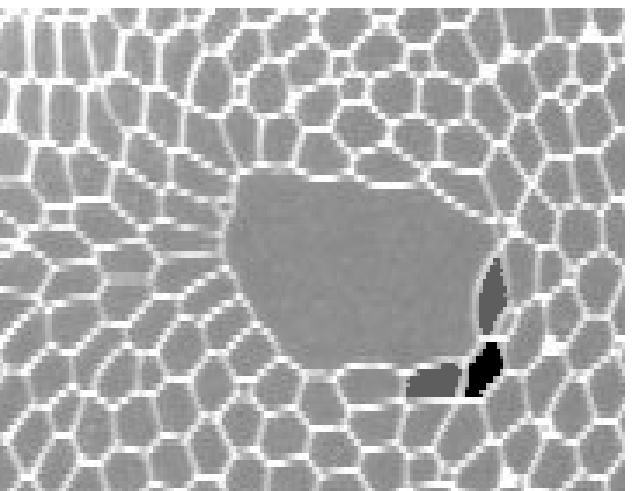}
\caption{\it Details of the raw images obtained from the camera. The foam, lit laterally by a circular neon tube, appears
 in white on black background.  The mean flow is oriented to the right and
the two images are separated by 120 ms. Small bubbles typical size is d=5mm.  The large
bubble is migrating through the smallest ones due to  T1 events : one of these elementary plastic transformations
occurs between images (a) and (b) and  the involved bubbles are underlined with a darker grey level.}
\label{T1}
\end{figure}

The foam is lit laterally by a circular neon tube of diameter 0.4 m, put horizontally just below the cell on a black board. 
The Plateau border network reflects the light at $90^o$ and thus appears from above in white on a black background
 (see Fig.\ref{T1}).
Only the Plateau borders perpendicular to the incident light lead to a good contrast and the isotropy of the light is thus 
crucial.  
The camera is placed above the middle of the cell, and the large bubble shape
 has then time enough to relax spontaneously from its initial arbitrary shape before reaching the recording zone.
We record 25 images per second with a Pulnix TM6CN camera, with a resolution
of $570 \times 760$ pixels. Images are processed using a commercial
software named Visilog to isolate each bubble and determine its
area $a$ and center of mass. The polygon area $a$ is related to the bubble volume $V$ through the relation $V=ha$. This 2D language will be used without precision in the following. The images treatment consists in the following operations : we subtract the background to remove the large scale heterogeneities of lightning and we get binary images by
thresholding (liquid is white and gas black). Then a white pixel dilatation improves the bubble separation and the continuous white network thus obtained is skeletized to get finally a one pixel width continuous frontier around each black bubble. The last step is to record  the area 
and the center of mass of  these connex domains of back pixels. 
Bubble tracking between images n and n+1 is then performed. A bubble k in image n is identified with the bubble
k' in image n+1 if its center of mass (computed in image n) belongs to k' in image n+1. We checked manually the pairing 
method on few images and found that the error ratio was less than few percents, so the individual bubble trajectories are determined
with a good precision. 

 The dissipation in the foam is mainly due to the viscous forces between the
 glass plates and the Plateau borders. The viscous force  per unit Plateau border length is predicted by the lubrication theory 
 to be ${\bf f}_v \sim -\gamma Ca^\alpha {\bf u}_x$, with $Ca = \eta v_0/\gamma$ the capillary number, $v_0 {\bf u}_x$ 
 the Plateau border velocity
  and $\alpha$ an exponent which is  $2/3$
 in the limit of  wet foam and mobile surfactants \cite{bretherton61,hirasaki85} and that varies roughly between 
0.5 and 0.7 in the general case \cite{denkov05}. For the present foaming solution and in the velocity range investigated, we measured $\alpha = 0.5 \pm 0.05$ with the technique detailed in \cite{cantat04}. These viscous forces are simply balanced by the pressure field, 
 which scales,  in a monodisperse foam of typical bubble diameter $d$, as 
\begin{equation}
P(x) \sim - x \, \gamma Ca^\alpha /(h d) \; .
\label{Pdex}
\end{equation}
The total pressure gap $\Delta P$ between upstream and downstream was measured with a pressure captor put in the foaming solution. 
 For a flow velocity $v_0=3.2 \, 10^{-2} m/s$
 and a small bubble size $d = 5 \, 10^{-3} m$, we measured $\Delta P= 2800$ Pa,
  leading the typical viscous force per unit length of Plateau border $f_v = h \Delta P d /(2L) \sim  10^{-2} N/m$, in good agreement with previously obtained data \cite{cantat04}.

\section{Qualitative behavior} 

\begin{figure}[h]
\centering
\includegraphics[width=6.5cm,angle=-90]{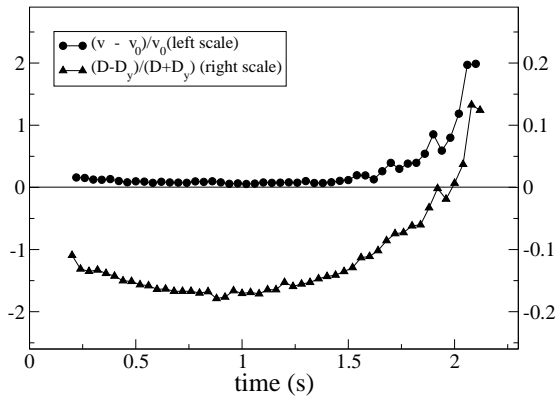}
\caption{\it The large bubble $v$ is compared to the average flow velocity $v_0$, and correlated to the large bubble shape (the LB diameters in the x and y direction are respectively $D$ and $D_y$).
An elongated shape oriented across the flow (large $D_y$) gives rise to a stationary plug flow $(0 < t < 1.5)$. This shape may
spontaneously destabilize $(1.5 < t < 2)$ : a tip grows downstream in the flow direction, similarly to a viscous digitation. The tip finally
contains the whole LB which is then oriented along the flow $(t > 2)$. This transformation coincides with an acceleration
of the large bubble.
}
\label{correlation}
\end{figure}
\begin{figure}
\centering
\includegraphics[width=6.5cm,angle=-90]{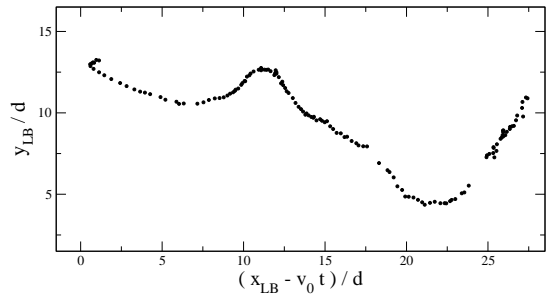}
\caption{\it Large bubble trajectory through the network of small bubbles. The position $(x_{LB}, y_{LB})$ of the large bubble is measured in the frame 
moving with the mean velocity of the foam $v_0 {\bf u}_x$, in unit of small bubble diameter $d$. The foam velocity is larger than the velocity threshold and the large bubble moves in the $x$ direction with a velocity fluctuating in amplitude and direction. Each 
dot corresponds to an image, separated by 40 ms.
  }
\label{fluctuation}
\end{figure}

The flow is obtained with a controlled foam flux. At the velocity range we investigated, 
 the first bubble layer on both sides slips on the wall at the mean velocity.
In monodisperse foam,  topological transformations
 are thus very rare  and the foam structure remains almost unchanged during the flow.
The instantaneous velocity distribution is well fitted
 by a Gauss distribution with a standard deviation of the order of $\pm 5\%$, due to the pixelisation. 
In contrast, in the presence of a large bubble, the plug flow becomes unstable above a given foam velocity.
The large bubble velocity is then larger than the mean flow velocity $v_0$, with relatively large amplitude fluctuations and an orientation that
may vary between $\pm 30^o$  with respect to the mean flow direction (see Fig. \ref{fluctuation}).
The large bubble shape was characterized by its diameters in the direction of the flow $D(t)$ and in the transverse direction $D_y(t)$. These two parameters depend on time, in contrast with the LB area
which is constant. The LB velocity is strongly correlated to the bubble elongation $(D-D_y)/(D+D_y)$ that varies roughly between -1/2 and 1/2. 
The large bubble alternates periods during which it moves much faster than typical bubbles, with a shape
elongated in the direction of its own relative motion, and periods during which it moves at the average speed with
an elongated shape perpendicular to the mean flow (see Fig. \ref{correlation}).

The large bubble migration occurs without films breakages. Small bubbles are separated in front of the large bubble and are reconnected at the rear. These so-called T1 events, depicted on Fig. \ref{T1}, are mainly localized very close to the large bubble, but many structure reorganisations are also induced much further, in the whole camera field.
The velocity field around the large bubble may exhibit very different behaviors. Indeed, the presence of a wake
of small bubbles behind the large one is intermittent. Comparison with foams flowing around an obstacle of fixed shape would thus be very interesting to perform\cite{dollet05}.

\section{Theoretical predictions}

This instability was already studied numerically and theoretically in \cite{cantat03a, cantat05a}. 
We summarized here the theoretical predictions for the large bubble velocity to improve the article self-consistency. 
The scaling behavior of the threshold is obtained from 
the expression of the  various forces acting on the large bubble. 
 From the pressure field given in eq. \ref{Pdex},  we obtain the resulting pressure force
on LB,  
\begin{equation}F_d \sim D^2 (\gamma/d)\, Ca^\alpha \; ,
\label{Fd}
\end{equation}
 which  is the driving force for the LB migration. For sake of simplicity, we only retain in the model a single diameter  D for the large bubble.
 Below the threshold, 
the LB equilibrium thus results from a competition between the pressure field pushing LB downstream (the driving force $F_d$) and the surface tension force 
of the soap films network that try to keep LB at its initial position in the foam frame. At small deformations, this force $F_e$, due to the fact that any modification of the foam structure increases the total amount of interfaces (and then of energy), is well predicted by a purely elastic model\cite{weaire94}. This  leads to
\begin{equation}
F_e \sim \gamma h X /d \, ,
\label{Fe}
\end{equation}
with $\gamma/d$ the order of magnitude of the foam shear modulus \cite{weaire} and X the LB displacement with respect to its equilibrium position.
This expression is obtained 
from a continuous incompressible 2D elastic medium at rest at infinity. In this case, the force exerted on a circular obstacle  scales indeed as the obstacle displacement time the 
material shear modulus (up to disregarded logarithmic corrections in the obstacle size).  

The local elastic stress $\sigma$ is related to $F_e$ by $F_e = \sigma L_{geo} h$, with $L_{geo} h$ the typical area on which the stress is important around the front or the rear of
LB. At the instability threshold, this local stress reaches the plastic threshold of the foam, scaling as $\gamma/d$ \cite{weaire94}. 
The force balance on the large bubble is then, from eq. \ref{Fd},
$ D^2 (\gamma/d) \, Ca^\alpha \sim \gamma L_{geo} h/d$ or, equivalently, with $v_{th}$ the velocity threshold,
\begin{equation}
\eta v_{th}/\gamma  \sim (h L_{geo}/D^2)^{1/\alpha} \; .
\label{eqseuil}
\end{equation}

Above this threshold, the large bubble begins to migrate, with a mean velocity v modelised by the
following process. 
The instantaneous LB  velocity v(t) decreases during the elastic loading of the foam until the 
plastic threshold is reached. Then T1 events occur and a new cycle begins. 
A full stress relaxation is assumed after each plastic events and the elastic force is zero.
Between two plastic events, the elastic force is obtained from eq. \ref{Fe},
$F_e = \gamma h /d \int_0^t (v(t')-v_0) dt'$. The integral is  the LB displacement X with respect to the small bubble network, since the last plastic relaxation, occurring at t=0. 
 The force balance is then
 \begin{equation}
 {\gamma h \over d} \int_0^t (v(t')-v_0) dt' -  {D^2 \gamma\over d} \left(\eta v_0 \over \gamma\right)^\alpha + D \gamma
   \left(\eta v(t) \over \gamma\right)^\alpha =0 \; ,
 \label{eqdiff}
\end{equation}
the first term being the elastic force given above, the second one the driving force (eq.\ref{Fd}) and the third one the excess of viscous 
force exerted around the large bubble that moves faster than the mean flow. 
This integral equation has been solved analytically in \cite{cantat05a}. In the simple case $\alpha =1$ and $D \gg d$ we obtain
for the LB mean velocity v the  expression given below. It differs only by few percents from the general expression, for reasonable values of $\alpha$
 and $D/d$, and will thus be used, for sake of simplicity (see Fig. \ref{vgb}). 
\begin{equation}
{v-v_0 \over v_0 {D \over d}} \sim {- v_{th} / v_0 \over \mbox{ln} \left( 1- {v_{th} \over v_0}\right )} \, .
\label{anal_approx}
\end{equation}
When $v_0 \gg v_{th}$, the expression becomes  asymptotically  $(v-v_0) / (v_0 D/d) = 1$ or equivalently 
$v = v_0 D/d$ with $D \gg d$. 
These equations \ref{eqseuil} and \ref{anal_approx} are used to rescale the experimental data.

\section{Experimental results}

The aim of the present paper is to determine experimentally the  relation,  at each time,  between the LB velocity in the flow direction, denoted by $v(t)$, the mean flow velocity
 $v_0$ and the large bubble size characterised by its diameter in the x direction $D(t)$, which appeared to be the pertinent size parameter. 

\begin{figure}[h]
\centering
\includegraphics[width=6.5cm,angle=-90]{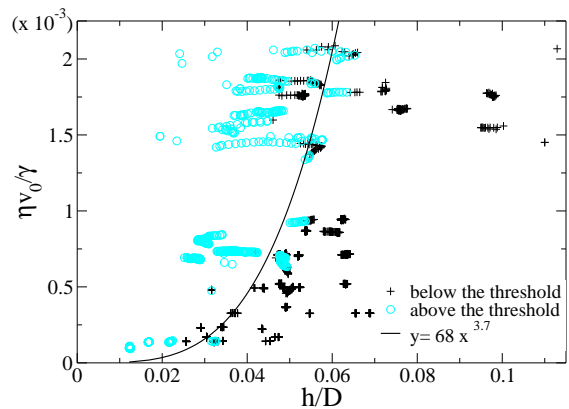}
\caption{\it Phase diagram of the large bubble instability. The control parameters are the capillary number built with the mean flow velocity $Ca=\eta v_0/\gamma$ and the LB diameter D adimensioned by the gap h between the plates. 
Each point of the graph corresponds to an image, on which the large bubble is migrating (circle) or not (cross). 
 Despite large fluctuations, a frontier between the stable and unstable domains
in the $(Ca, h/D)$ plane can be determined. 
The full line is the best fit of this frontier in power law,  $a\, x^b$ with a and b adjustable.
}
\label{seuil}
\end{figure}

We analyzed 56 movies with a large bubble remaining 2 or 3 seconds in
the view field. All control parameters were kept constant, except for the LB area and the mean flow velocity $v_0$.
The measured values of $v(t)$ and $D(t)$ were  averaged over 5 images.
When  $v(t) > 1.1 \, v_0$, a flow is considered as being above the
threshold, otherwise it is below.
All experimental points are represented in the plane $(h/D , \eta v_0/\gamma)$ on Fig.\ref{seuil}.
Despite
relatively large fluctuations, two distinct domains clearly appear on this phase diagram, corresponding to the two states, namely
above or below the threshold
(only the parameter ranges near the threshold were investigated).

To compare the experimental data presented on this figure \ref{seuil} with the theoretical prediction given by eq. \ref{eqseuil}, an adjustment of the experimental frontier has been performed by minimization of the distances $d_{front}$
 between the graph of a function and the experimental points located on the wrong side of this frontier.
 A least square method has been used with the set of functions
$f(x)=a\, x^b$,  $a$ and $b$ being free parameters.
We obtained thus the relation
\begin{equation}
\eta v_{th}/\gamma = 68 \, (h/D)^{3.7} \, . 
\label{eqseuilfit}
\end{equation}
The mean value of $d_{front}$ increases of $10\%$ if the exponent varies of $\pm 0.5$, which is the order of the precision 
on the exponent. This very crude estimation is nevertheless sufficient to discriminate between the two possible scaling for $L_{geo}$.
Indeed, $L_{geo}=D$ would correspond to an exponent $2$ in  eq.\ref{eqseuilfit} (with $\alpha = 0.5$), whereas $L_{geo}=d$ leads to the relation 
\begin{equation}
\eta v_{th}/\gamma  \sim (d/h)^2\, (h/D)^4
\label{seuil4}
\end{equation}
which is in  better  agreement with the relation \ref{eqseuilfit}. More experimental data would be needed to confirm the value of this exponent, but $h$ can not be easily varied and $d$ is confined to the small value range $h<d \ll D \ll l$ with $l$ the channel width. Nevertheless it tends to prove that $L_{geo} \sim d$ and further modeling will therefore have to go beyond the continuum medium approximation, in which the bubble size $d$ becomes irrelevant,
and to take explicitly into account its discrete nature, at least in the high stress regions
near the large bubble.

\begin{figure}
\centering
\includegraphics[width=6.5cm,angle=-90]{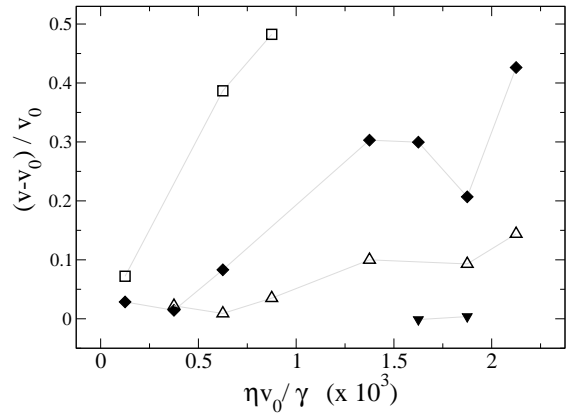}
\caption{\it Large bubble relative velocity $(v-v_0)/v_0$ as a function of the mean foam velocity $v_0$, for different large bubble size $D$ (the $h/D$ ratio are given on Fig.\ref{vgb} with the same symbol convention). Each point represents an average over few tens of measures. }
\label{vdehetD}
\end{figure}

\begin{figure}
\centering
\includegraphics[width=6.5cm,angle=-90]{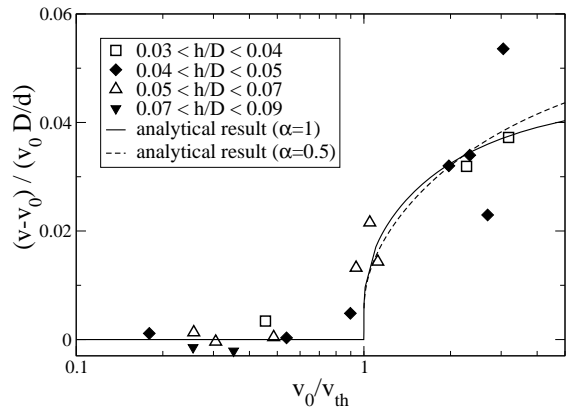}
\caption{\it Large bubble relative velocity in rescaled units (same data as in Fig. \ref{vdehetD}). The mean flow velocity threshold $v_{th}(D)$ used to rescale the experimental data,  is obtained from eq.\ref{eqseuilfit}. 
The first analytical result (full line) is obtained from eq. 
\ref{anal_approx}, with the adjustable prefactor 0.045. Numerical simulations in vertex model have been performed, leading to a good agreement with theory as well, with an adjustable prefactor of 0.07 \cite{cantat05a}. The second curve (dashed line) is the exact solution of eq. \ref{eqdiff}, using the experimental value of $\alpha$ ($\alpha=0.5$),
with an adjustable prefactor (see \cite{cantat05a}). 
  }
\label{vgb}
\end{figure}

Above the threshold, the large bubble migrates through the foam with
a velocity $v(t) > v_0$. Experimentally, the foam disorder induces large fluctuations of $v$, 
depicted on  Fig. \ref{fluctuation}, which have been averaged out. 
The averages were performed over
small bins of size $\delta (\eta v_0/\gamma) = 0.25 \, 10^{-3}$
and $\delta (h/D) = 0.01 \mbox{ or } 0.02$, leading to the graphs presented on Fig. \ref{vdehetD}. 
As expected, the large bubble velocity, in the frame of the foam, is vanishing at small foam velocity and/or small LB size. 
It can reach $0.5 v_0$ ({\it i.e} $1.5 v_0$ in the laboratory frame) for the largest foam flux explored. 
At larger velocities, images are not recorded fast enough by our camera to extract quantitative results. New physical processes, like films breakages, are involved and
modify the dynamical behavior.
For very large bubbles, the shape remains no more convex. The description of the phenomena in term
of a Saffman Taylor instability is then probably more appropriate \cite{park94,lindner00}.

The experimental data shown on Fig. \ref{vdehetD} were rescaled  according to
the theoretical prediction eq. \ref{anal_approx}.
The value used for the velocity threshold is the experimental fit given by eq. \ref{eqseuilfit}.
We obtain in that way a good superposition of curves for the various $h/D$ values, as well as an agreement with
the theory (see Fig.4).

The problem of the orientation of the LB velocity remains open. Local crystallization is presumably important, as
shown numerically. By contrast, disorder induces randomized plastic thresholds in the foam  and  thus creates most favorable
paths for the LB migration
that may deviate from a straight line. A precise analysis of the coupling between the local foam structure, the T1 localization
and the LB shape and orientation might help to clarify these questions.

\section{Large bubbles segregation in bidisperse foam}
\begin{figure}
\centering
\includegraphics[width=6.5cm]{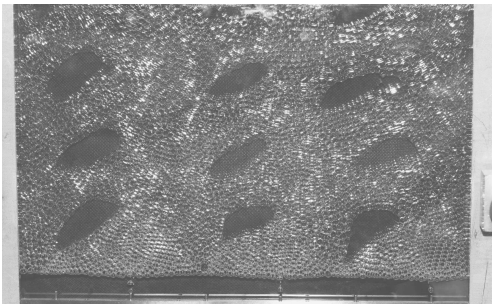}
\caption{\it Initial structure of the bidisperse foam. 
  }
\label{poly_init}
\end{figure}
\begin{figure}
\centering
\includegraphics[width=6.5cm]{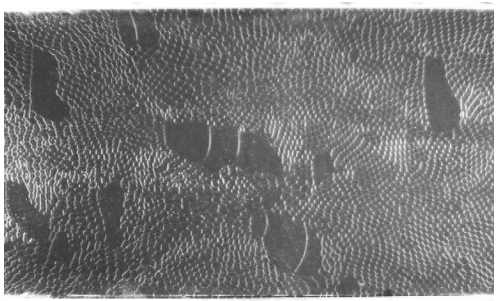}
\caption{\it Picture of the same foam as on Fig. \ref{poly_init}, taken few seconds later, 50 cm downstream in the cell.  The large bubbles have spontaneously organized in columns 
oriented along the flow.
  }
\label{poly_col}
\end{figure}
Polydisperse foam flows imply obviously interactions between bubbles of various 
sizes.
A  mean field approach, in which we would consider that each bubble migrates in an effective continuous medium and adjusts its velocity
as a function of its own size and of the local mean velocity and mean bubble size does not seem  appropriate. 
The flow appears indeed to be dominated by correlations between bubbles. 
In this last part, we point out the attractive interactions between a set of large bubbles, and their spontaneous organization 
in columns. 

\begin{figure}
\centering
\includegraphics[width=6.5cm,angle=-90]{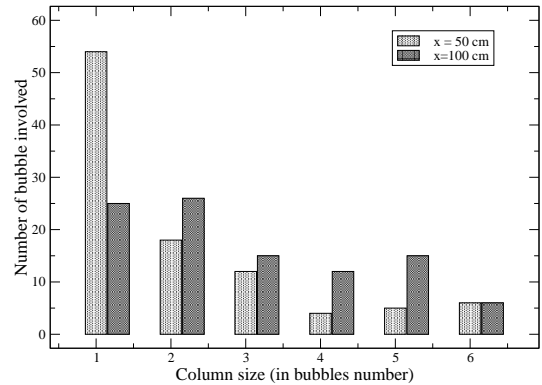}
\caption{\it Large bubbles repartition in the columns of various sizes, 
at two abscissa in the cell. Ten flows have been recorded, involving 99 large bubbles. 
The position $x=0$ correspond to the mean abscissa of the initial  regular large bubble network 
(see Fig. \ref{poly_init}). Each time a large bubble crosses the line $x=50 cm$ (resp. $x=100cm$), 
the size of the column in which this bubble migrates is recorded (size 1 corresponds to an isolated large bubble)
and the obtained distribution is plotted. The count is focused on the bubbles, which number is conserved, and the n-bubbles columns are thus counted n times.
  }
\label{poly_stat}
\end{figure}

A regular  network of approximately 10 large bubbles is initially produced in a monodisperse foam, at rest (Fig. \ref{poly_init}).
The middle of this network is used as abscissa reference ($x=0$).
Then the flow is turn on, at a velocity higher than the threshold, and the large bubbles begin to migrate through the foam. When two large bubbles meet, they migrate together. This aggregation leads progressively  to 
large bubbles domains of increasing size, organised in 
one bubble width columns, oriented along the flow (see Fig. \ref{poly_col}). 
 The length of the columns was measured at two points along the flow, for ten flows (99 bubbles). The number N(n) of bubbles involved in a n-bubbles column when crossing the abscissas $x=50$ or $x=100cm$ is plotted for each value of n on Fig. \ref{poly_stat} ($n=1$ for an isolated bubble). It represents equivalently the number of n-bubbles columns, with a weight n. 
 The initial distribution at $x=0$ is simply $N(1)=99$ and $N(i)=0$ for $i>1$. 
In order to follow every large bubble, we needed a large field of view. The image quality is therefore reduced 
and the analysis was done by hand.

The number of isolated bubbles decreases along the flow. The aggregation process may be due to attractive forces, 
or simply due to the spontaneous fluctuations of the large bubbles trajectories. Indeed, once in contact, two large bubbles 
never unbind, which leads to an effective attractive interaction, even without long range attractive forces. The orientation of the domains is probably 
governed by the same phenomenon as in viscous digitation.

\section{Conclusion}

In conclusion, the reported experiments evidence the crucial role of polydispersity in non quasi-static foam flows and open a whole field of investigations. We measured the relative
velocity of a single large bubble created in a monodisperse foam as a function of its  size and of the mean flow velocity.
The results are explained with non trivial scaling arguments. 
In a bidisperse foam, we show that the large bubbles are attracted to each other and organise themselves in columns oriented 
along the flow. 
A deeper analysis of the velocity fields around these large bubbles will allow for a better understanding of the nature of the attractive interaction. Strong
 spatial correlations are observed and a
mean field approach is probably not enough to explain the velocity field of a fully polydisperse foam.

The dissipation is of different nature in three-dimensional foams. Nevertheless, it is still localized
in the liquid phase, and the largest bubbles remain easier to deform or displace. Similar destabilizations
are thus expected to occur in non quasi-static 3D flows and might have important practical consequences.

{\bf Acknowledgments}:
We thank {\it Rennes M\'etropole} and CNRS for financial support, P. Chasles and S. Bourl\`es for technical
assistance and valuable advices and G. Le Ca\"er and B. Dollet for enlightening discussions.
%\bibliography{/home/serge/isabelle/bib/bib}
%\bibliographystyle{/home/serge/isabelle/bib/physrev}

\end{document}